\documentclass{aa}  
\usepackage{graphicx}
\usepackage{txfonts}
\usepackage{amssymb}
\usepackage{bm}
\usepackage{color}

\newcommand{\quotes}[1]{``#1''}

\begin{document}

\title{On the Identification of Coherent Structures in Space Plasmas: \\the Magnetic Helicity--PVI Method}


\author{
F. Pecora\inst{1}\fnmsep\thanks{francesco.pecora@unical.it},
S. Servidio\inst{1},
A. Greco\inst{1} \and
W. H. Matthaeus\inst{2}
}

\institute{
Dipartimento di Fisica, Universit\`a della Calabria, I-87036 Cosenza, Italy
\and
University of Delaware, Newark DE USA
}


 
  \abstract
   {Plasma turbulence can be viewed as a magnetic landscape populated by large and small scale coherent structures. In this complex network, large helical magnetic tubes might be separated by small scale magnetic reconnection events (current sheets). However, the identification of these magnetic structures in a continuous stream of data has always been a challenging task.}
   {Here we present a method that is able to characterize both the large and small scale structures of the turbulent solar wind, based on the combined use of a filtered magnetic helicity ($H_m$) and the Partial Variance of Increments (PVI).}
   {This simple, single-spacecraft technique, has been validated first via direct numerical simulations of plasma turbulence and then applied to data from the Parker Solar Probe (PSP) mission.}
   {This novel analysis, combining $H_m$~\&~PVI methods,
    reveals that a large number of flux tubes populate the solar wind and continuously merge in contact regions where magnetic reconnection and particle acceleration
   may occur. 
   }
   {}


\authorrunning{F. Pecora et al.}
\titlerunning{The Magnetic Helicity--PVI Method}

\maketitle
%


\section{Introduction}

The heliospheric plasma is embedded into a turbulent magnetic field originating from the Sun. Its rotating motion, together with fully developed turbulence, small scale magnetic reconnection, and wave-like activity produce complex topological structures that propagate away, filling the solar wind \citep{Jokipii66,JokipiiParker69}.  
Of the many types and sizes of magnetic 
structures that can emerge, perhaps
the largest 
are the sporadically occurring 
magnetic clouds, or Interplanetary Coronal Mass Ejections 
(ICMEs) 
that may extend more than an AU \citep{Burlaga81magnetic, Bothmer98structure, Scolini2019observation}
while having significant impact on 
particle propagation and energization.
However
magnetic structures may exist over a very 
wide range of scales and exhibit diverse morphology
and plasma properties\citep{Borovsky2008flux}. 
On the other hand,
a familiar, and even common, feature of these solar wind structures is the helical winding of magnetic field lines. This class of 
magnetic structures, such as the flux ropes (or tubes), are believed to fill a large fraction of the interplanetary volume and might be sites of heating and particle energization \citep{pecora2019statistical}. This perspective 
is often described as a ``tangled spaghetti model'' and dates back to the 
earliest models of the interplanetary medium
\citep{MccrackenNess66}. Afterwards, this 
interpretation was extended to 
view the interplanetary magnetic field 
as an ensemble of flux tubes, bounded by directional discontinuities \citep{Burlaga69}. The spaghetti model has been revived and interpreted on a number of more recent occasions \citep{Bruno2001identifying, Borovsky2008flux}, including the association of boundaries with tangential discontinuities \citep{greco2009statistical}. Discontinuities have been described as approximations to trapping boundaries \citep{tessein2016local,seripienlert2010dropouts} that can confine energetic particles within flux ropes \citep{tooprakai2007temporary,pecora2018ion} or, in the case of a solar moss model, exclude particles from the flux rope cores \citep{kittinaradorn2009solar}.

Though their origin still remains uncertain, there is a reasonable consensus that flux ropes -- or current carrying flux tubes -- can form close to the Sun and propagate outwards, or they can be generated locally by nonlinear interaction in the solar wind, or both.  Despite the wide range of scales these structures span, they have peculiar common signatures that often are clearly identifiable. For example, they show a rotation of one magnetic field component, accompanied by large magnetic field magnitude and density lower than the surrounding, ambient solar wind. Furthermore, they tend to assume a cylindrical symmetry of the magnetic field about a central axis. One may generalize this scenario to anticipate a wider variety of adjacent flux tubes (ropes) that have contrasting temperatures, densities and magnetic field strengths, but with approximate transverse total pressure balance \citep{Borovsky2008flux}.

Previous studies used the aforementioned properties to find flux tubes (or flux ropes or filaments) in the solar wind beginning with very detailed approaches that examine a number of parameters \citep{MccrackenNess66,Burlaga69,Borovsky2008flux}. These approaches allow distinctions to be made, and classifications of different types of magnetic flux structures. For example, the magnetic fields can be fitted to a Lundquist model to identify relaxed force-free states in magnetic clouds \citep{Burlaga88}. Other techniques also enable identification of other classes of flux tubes or flux ropes. Smaller scale flux rope such as ``plasmoids'' associated with by-products of magnetic reconnection \citep{Matthaeus86} can be implicated in particle energization \citep{Ambrosiano88,Drake06,KhabarovaEA16}, and occur frequently in turbulence \citep{WanEA14}. Detection of these structures has been proposed based on cross helicity, residual energy, and magnetic helicity evaluated using a wavelet analysis \citep{zhao2020identification}. In the realm of more elaborate techniques, one may also identify flux ropes by specializing in the case of near-equilibrium quasi-two-dimensional (2D) flux tubes. Then the reconstruction methods based on Grad-Shafronov (GS) equilibrium \citep{Sonnerup96,hu2002reconstruction} provide a pathway to visualize a 2D map of the flux tube cross section. It is important to bear in mind that the application of a technique such as GS-reconstruction  requires special conditions -- in this case, the total pressure must be, at least approximately, a single-valued  function of the magnetic potential \citep{Sonnerup16,Hu18}. A ``good reconstruction'', and therefore a reasonable detection of a GS flux tube, will be possible only when such auxiliary conditions are fulfilled \citep{chen2020small}.

In a previous work \citep{pecora2019single}, a GS method has been employed in conjunction with the Partial Variance of Increments (PVI) method \citep{greco2009statistical}, to identify near-equilibrium flux tubes and  nearby discontinuities. The clear result was that magnetic discontinuities are often found to populate both peripheral boundaries of GS flux tubes, as well as, in some cases, internal boundaries within the flux tubes. In these recent studies, one begins to find verification of the original conjectures that the interplanetary magnetic field consists of filamentary tubes bounded by discontinuities  \citep{MccrackenNess66,Burlaga69}, while recent advances provide much more detail to this picture \citep{KhabarovaEA16}. At this point, it may be useful to distinguish between identification methods that are more complex and involve specialized assumptions, such as the GS methodologies and the quantitative interpretation or fitting to Lundquist states. These more elaborate methods, which one might call reconstruction methods, typically provide more advanced information about the identified structures when they work - but they do not always work. 
While reconstruction methods can provide more information, detection methods can be more versatile and easier to implement. Reconstruction methods such as GS also can benefit from a preliminary identification step. It is on this point we focus the present study.

In the present paper, we construct a new method that is able to characterize both the large and small scale coherent structures of the turbulent solar wind, based the combined use of the magnetic helicity and the PVI. In some sense this approach can be viewed as a simplification and adaptation of the method employed by \cite{zhao2020identification}, a major difference being that, instead of passing over discontinuities, we employ PVI to detect them as potential boundaries. We exploit the helical nature of flux tubes to suggest a relatively straightforward alternative to the assumptions of two-dimensionality and equilibrium conditions. The proposed simple technique identifies a certain class of self-organized magnetic structures, with a minimum of assumptions, and no claim of exhaustive identification.

The paper is organized as follows. In Section~\ref{sec:HmPVI} we present the method, based in large scale filtered magnetic helicity and small scale gradients. We test the technique in Section~\ref{sec:simu}, by using direct numerical simulations of 2.5D compressible magnetohydrodynamics. In Section~\ref{sec:psp} we apply our technique to space data, by analyzing PSP dataset. Finally, in the last Section, we present our discussion and conclusions.

\section{Local Magnetic Helicity and PVI Methods}
\label{sec:HmPVI}
The approach is based on the assumption that magnetic flux tubes carry a finite amount of current density along their magnetic axis. These flux ropes are necessarily characterized by helical magnetic field lines near their magnetic axis, as a consequence of the Ampere's Law, in the case in which there is a non-null parallel (to the current) magnetic field component. These structures are inherently 2.5D, with spatial gradients that develop mainly in the 2D plane perpendicular to the current and a net magnetic field component along the current axis.
Envisioning (and simplifying) turbulence as an ensemble of quasi-parallel, large-scale flux tubes, those with the same polarity are often bounded by steep gradients such as small-scale, tangential discontinuities. In anisotropic turbulence, these represent regions of dynamical interactions between adjacent tubes, often observed in simulations [e.g., \citep{MattMont80,servidio2009magnetic}].  When present, these boundaries can be readily identified by a method such as PVI. The novel feature employed here is to exploit the frequently-occurring helical flux tubes, via a large scale analysis method based on the helicity, associated to a small scale PVI technique that identifies the reconnecting boundaries of such flux tubes.

The starting point is an ideal rugged invariant of MHD turbulence (in the absence of a mean magnetic field), namely the magnetic helicity $H_m = \langle {\bm a}\cdot \bm{b} \rangle$, where $\bm a$ is the magnetic potential associated to field fluctuations $\bm b$, while $\langle \dots \rangle$ represents an average over a very large volume, or over the interior of an isolated system \citep{Woltjer58b,Taylor74,MattGold82a}. 
This invariant can be estimated also by single-spacecraft, 1D measurements, as the out-of-diagonal part of the autocorrelation tensor \citep{matthaeus1982evaluation}:
\begin{equation}
    H_m = \int_\infty^0 d{s}_i ~ \epsilon_{ijk} ~ R_{jk}(\bm{\gamma}(s)).
    \label{eq:Hm}
\end{equation}
Here $R_{jk}({\bm{\gamma}}) = \langle B_j({\bm r}) B_k({\bm r}+{\bm{\gamma}}) \rangle$ is the correlation tensor evaluated at vector spatial  lag $\bm{\gamma}$, and the fluctuations are assumed to be well described by  spatially homogeneous statistics, up to the second order correlations.   The line integral is evaluated along a specified curve parameterized as ${\bm{\gamma}}(s)$ from  a specified origin at ${\bm{\gamma}} = 0$ to infinity.  The differential line element along ${\bm{\gamma}}$ is $d{\bm s} = ds~d{\bm{\gamma}}(s)/ds$  where $s$ is the  scalar displacement along the curve and $d{\bm{\gamma}}(s)/ds$ is a unit vector tangent to the curve. 
  
While the principle for evaluating fluctuating helicity has  become well known, it most often is applied  using suitable Fourier transforms to the evaluation of the  reduced, one dimensional magnetic helicity spectrum. Extending the spectral approach,  helicity measurement has also been implemented using wavelet transforms \citep{farge1992wavelet,bruno1999solar,telloni2012wavelet,zhao2020identification}. The alternative approach, implemented here, is based on consideration of the real-space formulation (\ref{eq:Hm}). This defining equation may be arbitrarily decomposed as 
\begin{eqnarray}
    H_m &=& \int_\infty^\ell ds ~ \hat e_i ~ \epsilon_{ijk} ~ R_{jk}(s) 
    + \int_\ell^0 ds ~  \hat e_i ~ \epsilon_{ijk} ~ R_{jk}(s) =\\
    & = & H_m^+(\ell)  + H_m^-(\ell), 
    \label{eq:Hm2}
\end{eqnarray}
where integral is now specialized to the case integration path in a fixed direction $\bm{\hat e}$ with scalar lag $s$. The obvious interpretation is that $H_m^+(\ell)$ is the contribution to helicity from structures larger than $\ell$ while $H_m^-(\ell)$ is the contribution to helicity from structures smaller than $\ell$.  The method employed below is the direct evaluation of the special case
\begin{equation}
   H_m^- (\ell)  =   \int_\ell^0 ds ~ \hat{e}_i ~ \epsilon_{ijk} ~ R_{jk}(s)
   \label{eq:Hminus}
\end{equation} 
where integral is again in the direction $\bm{\hat e}$ with scalar lag $s$.
It is important to emphasize that this approach will provide a measurement of the helicity of the fluctuations that have spatial 
scales less than $\ell$, the principle assumption being that 
of spatially homogeneous turbulence. 

{\it Implementation.}
In the usual way, the above mathematical formulation requires a practical interpretation of the ensemble average, usually accomplished by averaging in space or time, relying on an ergodic theorem \citep{Panchev}. Assuming a single spacecraft measurement is available, and the measurement point is fixed in space, averaging is done in one Cartesian direction, using the Taylor hypothesis \citep{Jokipii73}. 
In this familiar approximation, a spatial lag $s$ is 
inferred by computing a convected distance in a given time lag, assuming no distortion during this time interval. 
Therefore, with 
solar wind speed ${\bm V}=V{ {\bm{\hat e}}}$, one approximates $s = - V\tau$, with $\tau$ the time lag.

With these assumptions, 
the magnetic helicity of the fluctuations, and other derived quantities such as its reduced 
one-dimensional spectrum,  
may be derived from interplanetary spacecraft data \citep{MattGold82a}. Here we propose a procedure 
to calculate a {\it local} estimate of Eq.~(\ref{eq:Hminus}).
To obtain an estimate of the requisite elements of the correlation matrix
at the point $x$,  we average the local correlator (symbolically, ``$b_2b_3' -b_3b_2'$'') over a region of width $w_0$ centered about $x$. To avoid effects of large fluctuations at the edges of the interval of data 
that is investigated, a window is employed to smoothly let the estimates to zero at the edges, a procedure familiar in correlation analysis \citep{MattGold82a}. Specifically, in the first step, the raw helicity is estimated as 
\begin{equation}
C(x, l) = \frac{1}{w_0} \int_{x-\frac{w_0}{2}}^{x+\frac{w_0}{2}} \left[ b_2(\xi)b_3(\xi+l) -b_3(\xi)b_2(\xi+l)\right] d\xi.
    \label{hm1}
    \end{equation}
This is followed by windowing as    
\begin{equation}    
    H_m(x,\ell) = \int_0^\ell dl~C(x,l) h(l)
        \label{hm2}
\end{equation}
where $h(l) = \frac{1}{2}\left[ 1 + \cos\left(\frac{2\pi l}{w_0}\right) \right]$ is Hann window of the correlation function $C(x, l)$. Note that the interval of local integration $w_0$ is arbitrary, but we typically chose it as an order unity multiple of the scale $\ell$, such as $w_0=2\ell$. 
Hereafter we call this quantity computed in Eq.~(\ref{hm2}) as $H_m(x,\ell)$ or simply $H_m$ when it does not cause confusion. The above formulas convert directly to the time domain using the Taylor hypothesis. 

For determining the boundaries of a flux rope we use the PVI \citep{greco2008intermittent}, defined as
\begin{equation}
    \mbox{PVI}(s, \ell) = \frac{ | \Delta {\bm B}(s,\ell) | }{ \sqrt{ \langle | \Delta {\bm B}(s,\ell) |^2 \rangle } },
    \label{pvieq}
\end{equation}
where $\Delta {\bm B}(s,\ell) = {\bm B}(s+\ell) - {\bm B}(s)$ are the increments evaluated at scale $\ell$ and the averaging operation $\langle \dots \rangle $ is performed over a suitable interval. The function can be computed spatially in simulations or in magnetic field time series, by assuming the Taylor hypothesis. The technique has been strongly validated in different conditions \citep{greco2018partial}, identifying current sheets that spontaneously form in between magnetic islands in simulations \citep{greco2009statistical} and observations \citep{pecora2019single}. The technique, for very strong events, can detect local reconnection events in the turbulent solar wind \citep{OsmanEA14}.

Below we will implement a combined procedure that employs both the local real space magnetic helicity analysis and the PVI method. The purpose is to identify flux tubes and their boundaries, with examples given in simulation and spacecraft analysis using Parker Solar Probe data. 

\section{Analysis of Turbulence Simulations}
\label{sec:simu}

\begin{figure}
    \centering
    \includegraphics[width=.49\textwidth]{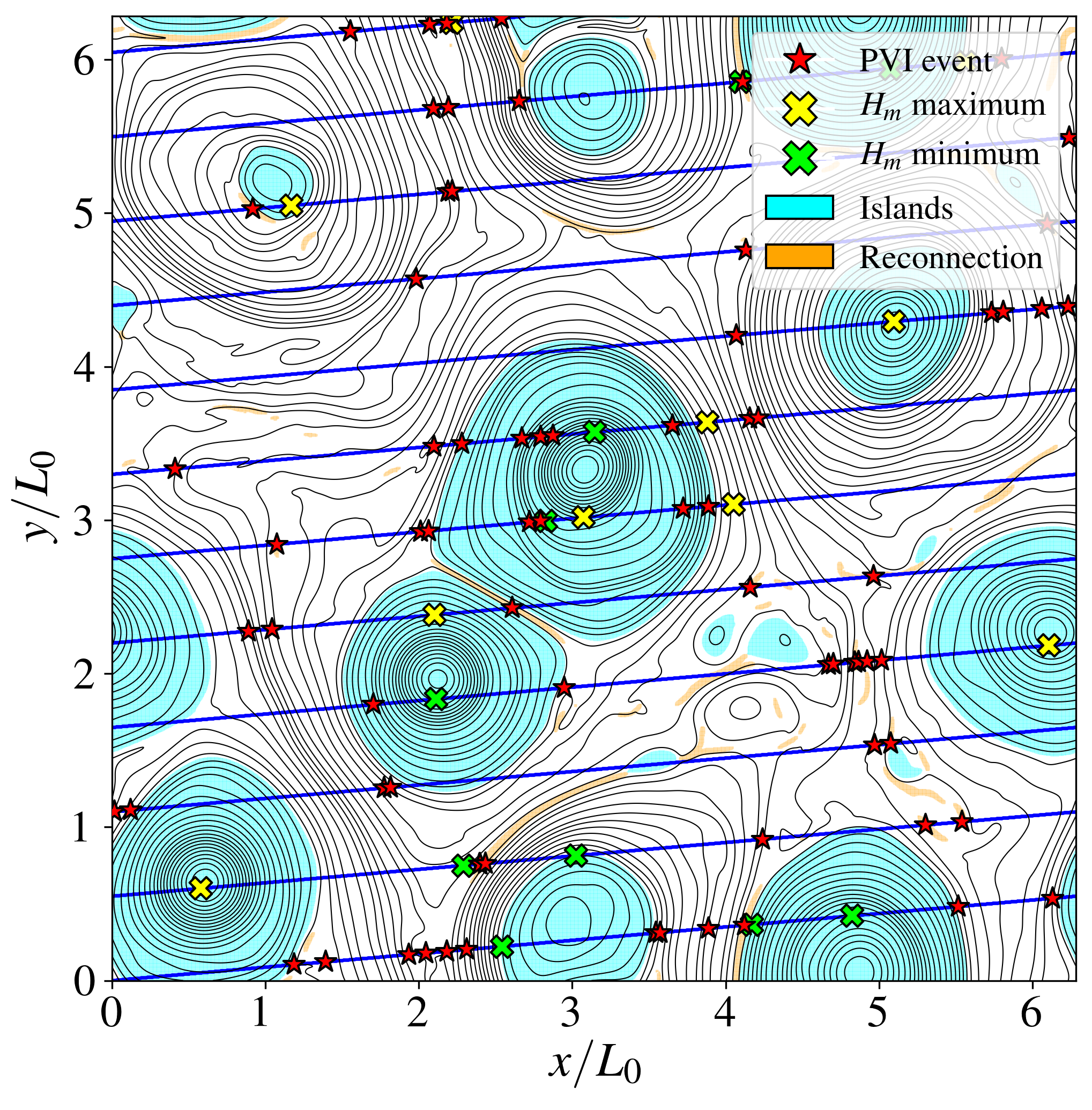}
    \caption{2D line contours of the magnetic potential $a$ (black solid), at the peak of the turbulent activity. The shaded areas in cyan and orange are magnetic islands and strong current sheets, respectively, as painted by the CA algorithm. The oblique (blue) lines represent the trajectory of a virtual satellite that sweeps through turbulence. (Yellow and green) crosses indicate maxima and minima of the local magnetic helicity, while (red) stars are strong PVI peaks.}
    \label{fig:whole}
\end{figure}

\begin{figure*}[hbtp!]
    \centering
    \includegraphics[width=.95\textwidth]{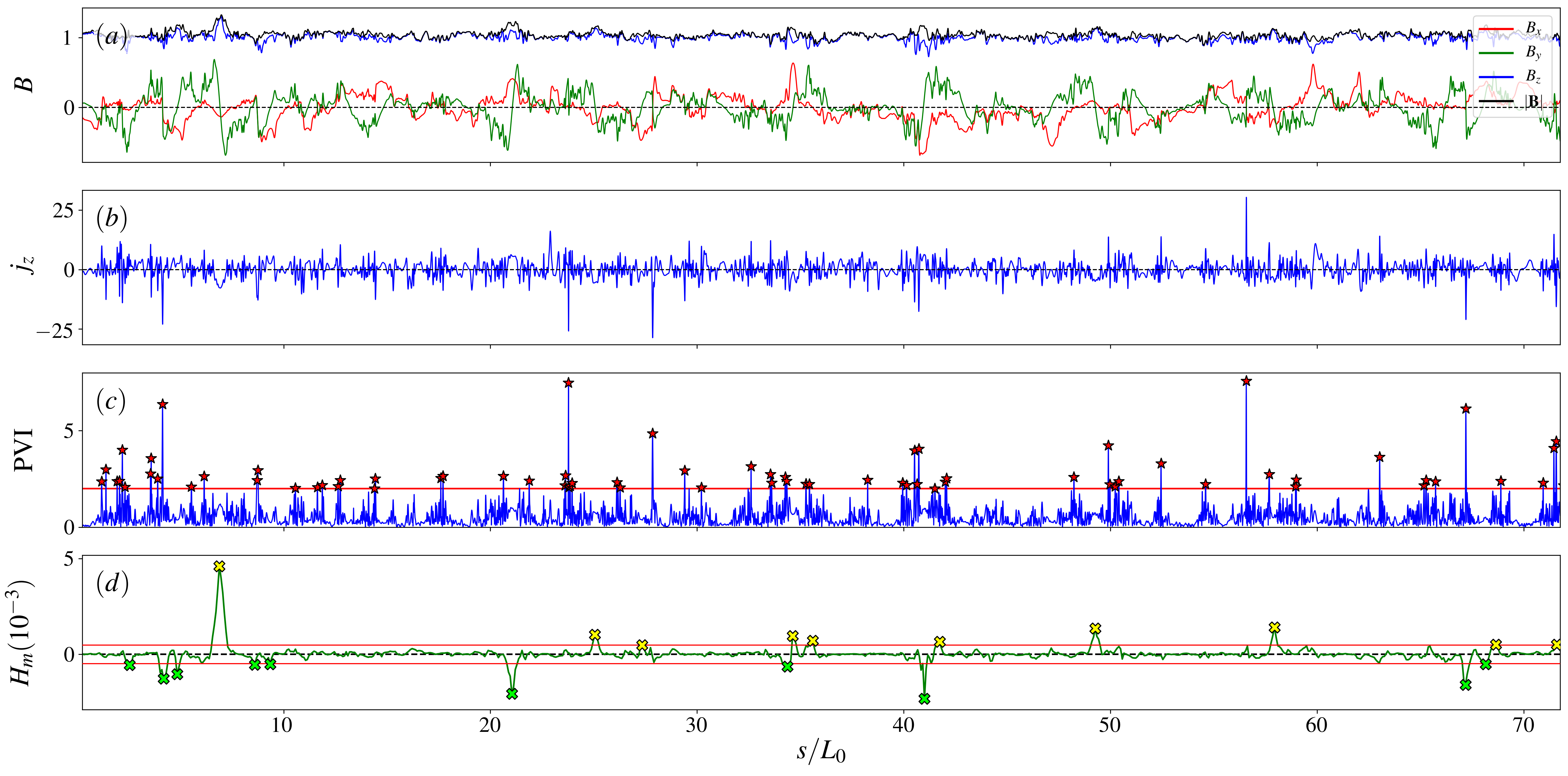}
    \caption{(a) Magnetic field components and magnitude; (b) Current density in the out-of-plane direction $j_z$; (c) Time series of PVI($s, \ell=\lambda_T/10$); (d) Filtered magnetic helicity evaluated at a correlation scale $H_m \left(s, \ell=\frac{\lambda_c}{2}\right)$. The horizontal (red) lines in (c) and (d) represent the thresholds of the method, where the symbols represent over-threshold peaks.}
    \label{fig:whole_HmPVI}
\end{figure*}

We test our novel technique by using direct numerical simulations of compressible Magnetohydrodynamics (MHD). We solve the equations in $2.5$D, in a square box of size $2\pi L_0$, with a resolution of $2048^2$ grid points. All the quantities are normalized to classical Alfv\'en units. The simulation has been performed in the $x$-$y$ plane and a mean magnetic field $B_0=1$ is present along the $z$ axis. The velocity and magnetic field fluctuations have all three Cartesian components. The code, based on very accurate pseudo-spectral method \citep{GottliebOrszag,GhoshEA93-cpc}, as described in Ref.~\citep{perri2017numerical},  makes use of a logarithmic density. In order to preserve the solenoidal condition of the magnetic field, the algorithm solves equations directly for the magnetic potential $a$ and parallel variance $b_z$, so that the total magnetic field that has been decomposed as ${\bm B} = B_z {\bm{\hat z}} + {\bm \nabla a}\times {\bm{\hat z}}$. Here $B_z = B_0+ b_z$ is the out of plane magnetic field, where ${\bm \nabla}= (\partial/\partial x, \partial/\partial y,0)$ is the in-plane gradient. The algorithm has been stabilized via hyperviscous dissipation to suppress very small scale, spurious numerical effects. For these simulations, in order to limit the effect of dissipation at very high $k$-vectors, we use a small, fourth-order hyperviscosity, with coefficients on the order of $10^{-8}$). This numerical tool has been extensively used in the past years with applications to space plasma turbulence \citep{VasconezEA15, matthaeus2015intermittency,perri2017numerical}.

The present simulation parameters are in the range of solar wind conditions, with plasma $\beta=1$ and fluctuation amplitude $\delta b/B_0=1/2$, where $\beta$ is the ratio of kinetic to magnetic pressures and $\delta b$ is the total r.m.s magnetic fluctuation amplitude. The initial fluctuations are chosen with random phases, for both magnetic and velocity field, in a shell of Fourier modes with $3\leq |{\bm k}|\leq 5$, where the components of wave vector are in units of $1/L_0$. The decaying MHD simulation quickly develops turbulence and small scale dissipative structures. The magnetic field power spectrum (not shown here) manifests a power-law typical of Kolmogorov turbulence, namely with a scaling $P(k)\propto k^{-5/3}$. The turbulent pattern is represented in Fig.~\ref{fig:whole}, where we show the 2D contour lines of constant magnetic potential $a$ (black solid lines), which are readily identified with the in-plane projection of the magnetic field lines. 
The typical features of 2D turbulence are evident, with large scale coherent structures and narrow discontinuous contact regions, where frequently one finds that reconnection is occurring \citep{servidio2009magnetic}. On the same figure, we report, as shaded areas, magnetic flux tubes and reconnecting current sheets, identified using a cellular automaton (CA) procedure that is described in Ref.~\citep{Servidio11}.  This CA is built on the topological properties of the magnetic potential. In particular, we first identify the critical points (maxima, minima and X-points) and then the algorithm propagates information away from these critical points, identifying the strongest, large scale islands (starting from the O-points) and the reconnection regions (starting from the X-points). The result of this procedure is a cellularization of turbulence, as clear from Fig.~\ref{fig:whole}. These islands retain a finite amount of magnetic helicity, due to non-zero parallel variances ($b_z$) that are frequently concentrated 
near the center of the flux ropes.

On this magnetic skeleton we test our 1D algorithm, based on the combination of the local magnetic helicity in Eq.s~(\ref{hm1})-(\ref{hm2}) and the PVI method in Eq.~(\ref{pvieq}). In order to test the method and to establish a direct comparison between the plasma simulation and the PSP data, we send a virtual spacecraft trough the periodic domain. Its trajectory is represented in Fig.~\ref{fig:whole} with oblique (blue) lines, which intersects both large scale helical structures (cyan) and small scales discontinuities (orange). In Fig.~\ref{fig:whole_HmPVI}-(a) we report the turbulent magnetic field, as observed along the virtual satellite trajectory. 

The 1D signals are shown over the entire trajectory along the oblique coordinate $s$ measured in units of $L_0$. In (b) we show the out of plane current $j_z$ which is very intermittent, indicating the presence of magnetic discontinuities. To identify these intermittent spots, by using the interpolated magnetic field, we computed the PVI signal, as described in Eq.~(\ref{pvieq}). We used very small increment lags, namely PVI$(s, \ell=\lambda_T/10)$, where $\lambda_T = \sqrt{ \delta b_{rms}^2 / j_{rms}^2 } $ is the magnetic Taylor microscale. At these lengths, the time series generated by the PVI method becomes a good surrogate for the current density, as is suggested by comparing panel (b) and panel (c) of the same figure [for more on this comparison, see \citet{greco2018partial}].

To complete the analysis, we computed the filtered magnetic helicity in Eq.s~(\ref{hm1})-(\ref{hm2}). First we rotated the magnetic field ${\bm b}$ from the Cartesian ($b_x, b_y, b_z$) frame to the trajectory coordinates ($b_1, b_2, b_3$), where $b_1$ is the component along the trajectory direction ${\bm {\hat e}}$, $b_3$ remains along $z$ and $b_2$ completes the right-handed frame. Second, from this rotated field we computed $H_m$ signal, at the (cumulative) scale $\ell=\lambda_c/2$. As it can be seen from the lower panel of Fig.~\ref{fig:whole_HmPVI}, there is a net offset between the large helical flux tubes and the PVI peaks.

\begin{figure}[hbtp!]
    \centering
    \includegraphics[width=.49\textwidth]{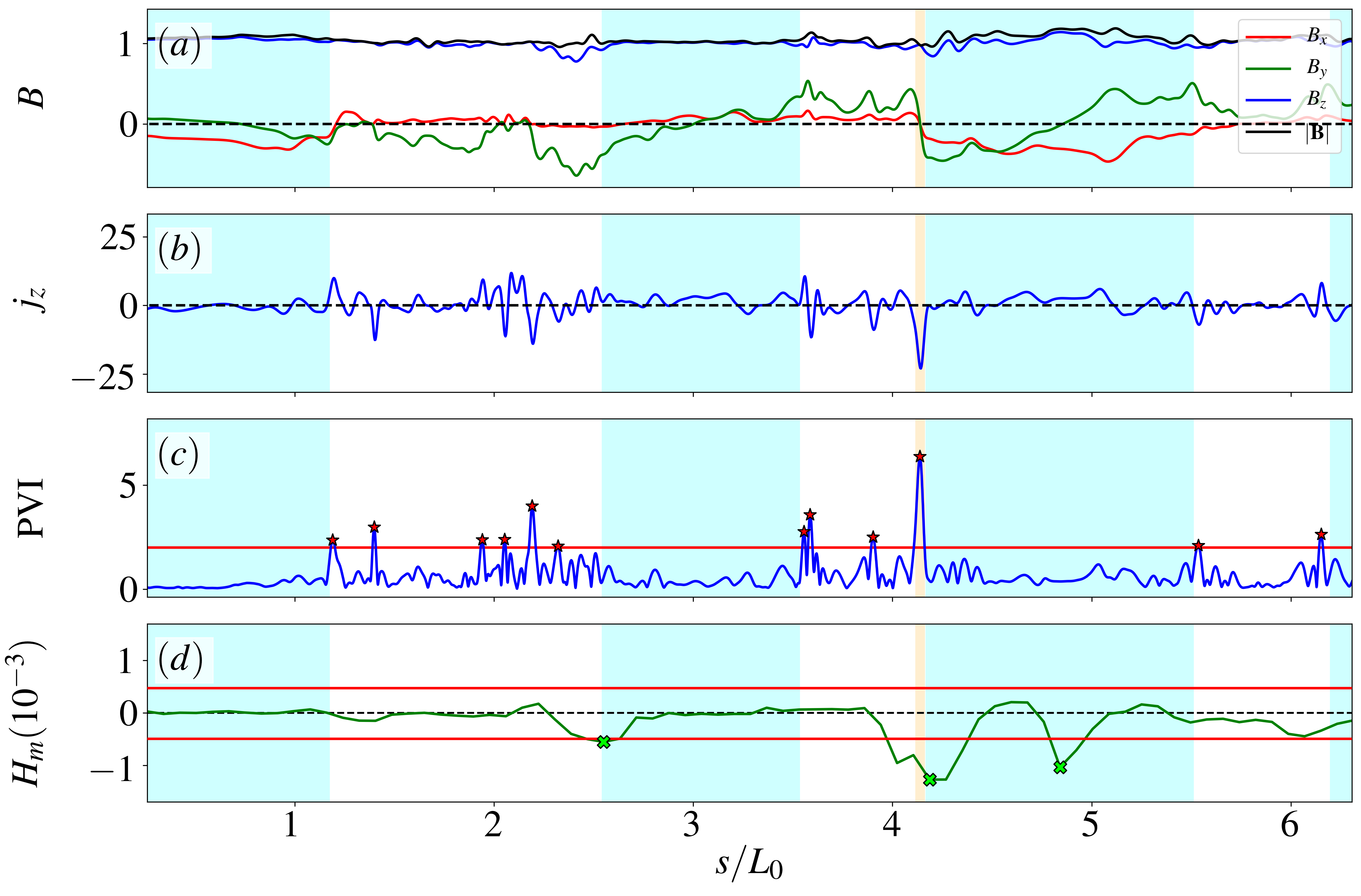}
    \caption{Zoom on magnetic field, current density, PVI and $H_m$ for the first segment of the trajectory where three peaks of helicity and twelve PVI events have been identified. Shaded cyan and orange areas represent the CA structures reported in Fig.~\ref{fig:whole}. The  $H_m$ peaks fall in PVI-quiet regions, in between consecutive, strong PVI clusters.}
    \label{fig:zoom_line1}
\end{figure}

By using both the surrogate $H_m$ measurement and the PVI signals, we established a threshold-based method in order to identify the most significant events. For the $H_m$ we identified as strong flux tubes the events with helicity values larger than one standard deviation of the $H_m$ distribution. For the PVI method, we chose a typical threshold of PVI$=2$. 
It has been shown that the probability distribution of the PVI statistic
derived from a non-Gaussian turbulent signal strongly deviates
from the probability density function of PVI computed from a Gaussian signal, for values of PVI greater than about 2. As PVI increases the recorded “events” are extremely likely to be associated with coherent structures and therefore inconsistent with a signal having random phases \citep{greco2018partial}.

The selected peaks are reported, for both $H_m$ and PVI, on panel (c) and (d) of Fig.~\ref{fig:whole_HmPVI}, respectively.  At this point, we have a list of selected events, namely the position of the possible flux ropes (peaks of the filtered magnetic helicity signal) and the reconnection events (peaks of the PVI signal). The position of these coherent structures is recorded on the full 2D map in Fig.~\ref{fig:whole}, and one may observe a very good qualitative agreement of these events with the magnetic potential and the CA-painting. Magnetic helicity peaks are well located inside helical islands, close to their cores. A few are located outside and coincide with PVI events, indicating the presence of complex structures in between islands, possibly due to reconnection-induced reorganization of magnetic field topology. On the other hand, red stars, the PVI events, are found at the boundaries of magnetic islands. This precise positioning of magnetic helicity and PVI peaks, suggests that the core of a magnetic island can be identified, with noteworthy precision, by a magnetic helicity extremum, and its boundaries coincide well with the closest PVI events on either side. The new method is able to identify the strongest helical flux tubes and the more intermittent magnetic structures, the latter likely to be reconnection events.

Fig.~\ref{fig:zoom_line1} shows a close up of the relevant quantities measured along the first segment of the synthetic trajectory. From this 1D information, the identification of magnetic islands is rather straightforward: the oscillations of the PVI signal tend to drop in magnitude near an extremum of local $H_m$, and the boundaries of the islands are well defined by a sharp increase of PVI. Moreover, inside the current cores, it is evident that the total current is smaller in general, but not zero, in view of Ampere's law. The net magnetic helicity in a flux rope is indicated by a non-zero component of the out of plane magnetic field fluctuation $b_z$. It is interesting to notice how well the $H_m$ peaks fall in a PVI-quiet region in between two strong PVI events.  The second $H_m$ peak is superimposed to a PVI event and is not associated with island-like structures as the map in Fig.~\ref{fig:whole_HmPVI} shows. 
\begin{figure}[hbtp!]
    \centering
    \includegraphics[width=.49\textwidth]{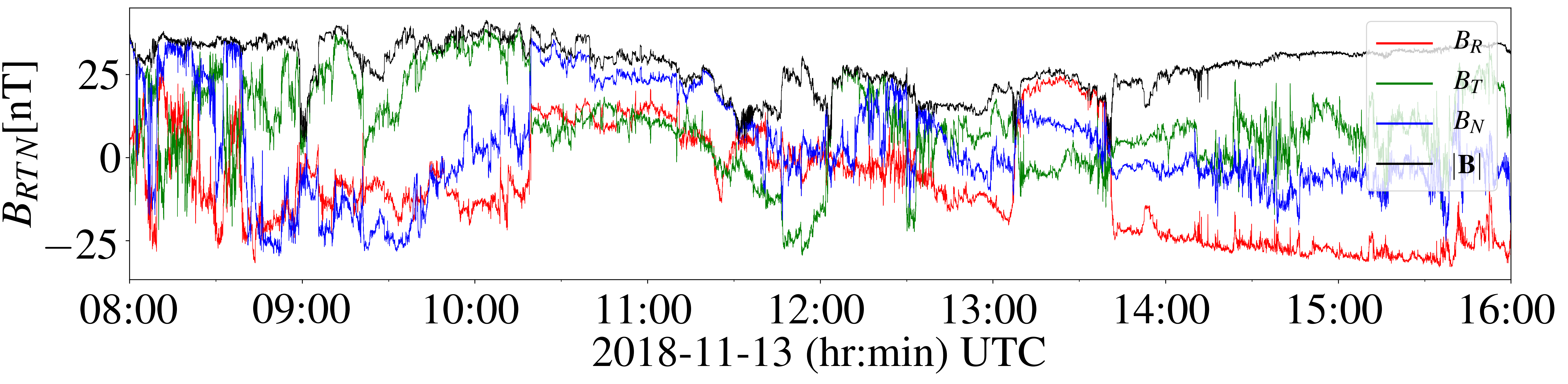}
    \caption{PSP FIELDS Fluxgate Magnetometer (MAG) data in RTN coordinates resampled at $1s$ on 2018 November 13 from 8:00 to 16:00 UTC. In this 8-hour interval, the correlation time $\tau_C \sim 25$ minutes.}
    \label{fig:PSP20181113}
\end{figure}
This can be due to turbulence and small scale structures that entangle magnetic field topology locally, resulting in both discontinuities and helical small regions.

\section{Analysis of Parker Solar Probe dataset}
\label{sec:psp}

We apply the  $H_m$--PVI technique to the fluxgate magnetic field data obtained by the PSP FIELDS instrument suite \citep{BaleEA16}. In particular, we analyze results obtained from the first perihelion \citep{BaleEA19} that will be further discussed in comparison with other identification techniques \citep{zhao2020identification, chen2020small}. The FIELDS magnetic data have been resampled from full resolution to 1-second cadence.  Moreover, the first encounter data has been divided into 8 hour-long subsets in order to contain several correlation times for each dataset. The correlation time $\tau_C$, in the spacecraft frame, is about 10-40 minutes at radial distances of $0.17-0.25$ AU \citep{ParasharEA20}. We analyzed several such intervals in the first encounter; however, to make close contact with above mentioned published works \citep{zhao2020identification, chen2020small}, we concentrate below on a particular interval -- 2018 November 13 from 8:00:00 to 16:00:00 UTC. Magnetic field time series, in the RTN coordinate system, for this interval is shown in Fig.~\ref{fig:PSP20181113}. In this interval, the average plasma $\beta \sim 1$ as reported by \citep{chhiber2020clustering,zhao2020identification,chen2020small}. %

\begin{figure*}[hbtp!]
    \centering
    \includegraphics[width=.95\textwidth]{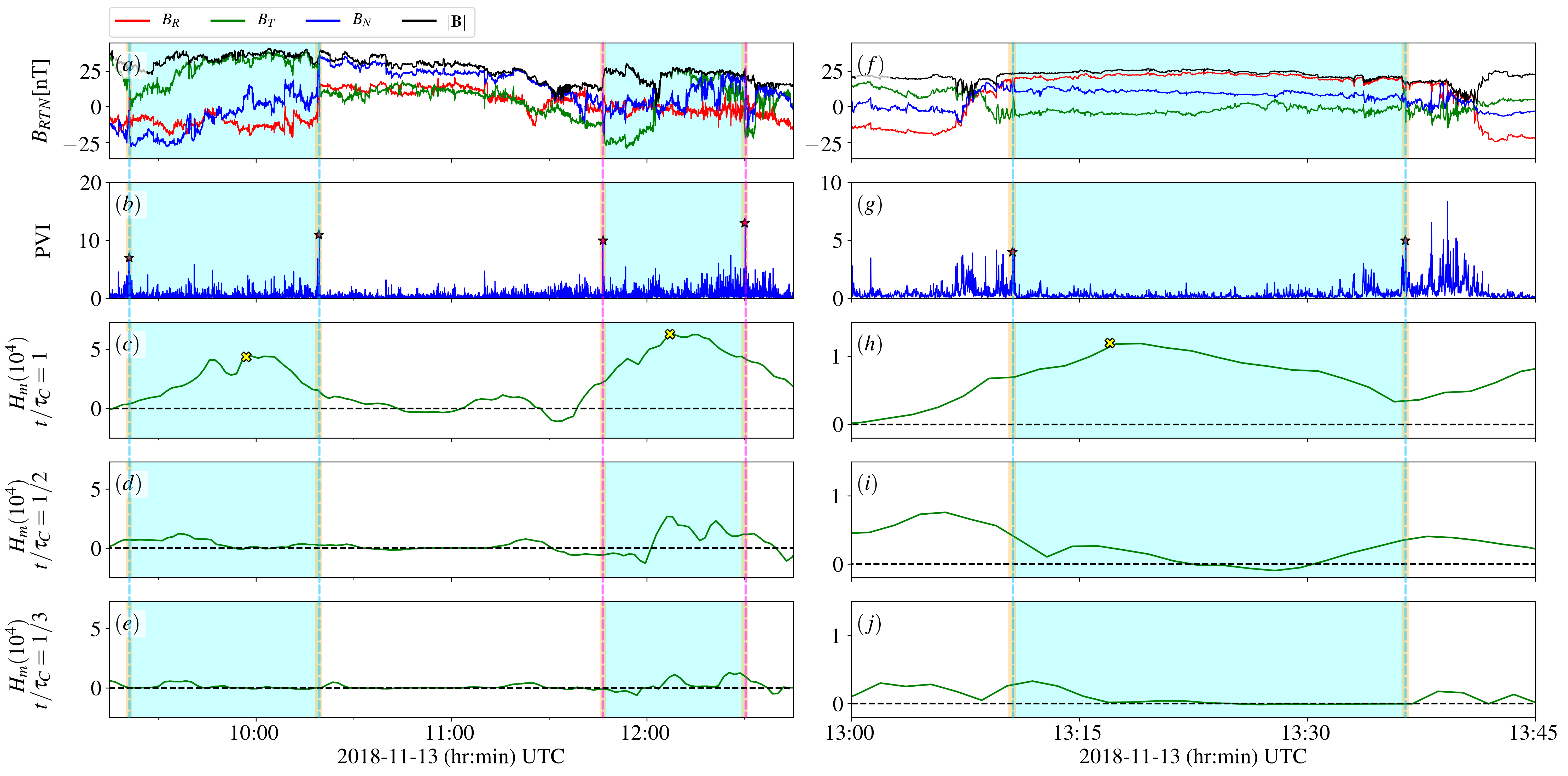}
    \caption{Two close-up of Fig.~\ref{fig:PSP20181113} from 9:15 to 12:45 (left panels) and from 13:00 to 13:45 (right panels). The figure reports the magnetic field (a),(f), the PVI signal (b),(g) and the magnetic helicity evaluated at one correlation time, (c),(h), $1/2$ correlation time (d),(i) and $1/3$ of the correlation time, (e),(j). It is interesting to notice that $H_m$ shape heavily depend on the chosen window, suggesting a multi-scale nature of helical structures. The vertical cyan and magenta lines highlight the position of strong PVI events edging high-helicity regions (cyan shaded regions). Left panels show two structures bounded by two strong PVI events each (cyan lines for the first and magenta lines for the second). At scales smaller than one correlation time, $H_m$ signal shows a fragmentation of the structure, highlighting sub features within the helical structure. Moreover, the two identified cores might be enclosed within a larger helical structure, possibly bounded by the leftmost cyan and rightmost magenta lines. This description is also consistent with the reconstruction performed by \citep{chen2020small} (Fig.~4) that shows a large island with two inner structures. On the other hand, right panels show a smaller structure at $t/\tau_C=1$, clearly bounded by PVI events, that has no internal features.}
    \label{fig:PSP20181113_zoom}
\end{figure*}
Fig.~\ref{fig:PSP20181113_zoom} shows analysis of two sub-intervals of the data shown in Fig.~\ref{fig:PSP20181113}, specifically from 9:15 to 12:45 (left panels) and from 13:00 to 13:45 (right panels). Each column shows stacked plots of the magnetic field time series, the PVI, and the local magnetic helicity, computed with different maximum lags. The largest lag was chosen to be one correlation time $\tau_C$ while the smallest was $t = 1/3 \tau_C$. Regions of high helicity have been shaded in cyan while edging PVI events in orange, for similarity with simulation. Moreover, pairs of PVI events which bound helical regions are also highlighted with dashed vertical lines. It is evident that the $H_m$ time series suggest a multi-scale nature of helical structures. We recall that the helicity diagnostic incorporates contributions from all scales smaller than the maximum lag.

Left panels of Fig.~(\ref{fig:PSP20181113_zoom}) show two helical structures bounded by two strong PVI events each (cyan lines for the first and magenta lines for the second). At scales smaller than one correlation time (panels~(d)~and~(e)), $H_m$ signal shows a fragmentation of the structures, highlighting smaller features within and near the larger helical structure. Moreover, the two identified cores (cyan shaded regions) might be enclosed within a larger helical structure, possibly bounded by the leftmost cyan and rightmost magenta lines. This description is also consistent with the reconstruction performed in Ref.~\citep{chen2020small} that shows a large island with two inner structures at about the same period. On the other hand, right panels show a single structure at $t/\tau_C=1$, in panel~(h) clearly bounded by PVI events. At smaller scales, $t/\tau_c = 1/2$ (panel~(i)) and $t/\tau_c = 1/3$ (panel~(j)), $H_m$ signals suggest the absence of relevant internal structures, envisioning a \quotes{pristine} flux rope.

\section{Discussion}
\label{sec:discussion}
Substantial progress has been made in recent years in identifying magnetic and plasma structures in the solar wind based on the flux tube paradigm \citep{Bruno2001identifying,Borovsky2008flux} and also on the analysis of discontinuities \citep{Neugebauer06,VasquezEA07-disconts}. Such studies build upon the concept that a substantial volume fraction of the solar wind magnetic field is organized into ``filaments'' or flux ropes \citep{MccrackenNess66}, that is current carrying flux tubes that maintain their integrity over some reasonable distance. Eventually, due to turbulence, a typical flux tube will become highly distorted and ``shredded'' over sufficiently large distances along the magnetic axis \citep{ServidioEA14-flux}. The idea that these tubes might act as conduits for solar energetic particles \citep{MccrackenNess66} 
has recently received theoretial and 
observational support \citep{Ruffolo03,TesseinEA13,tessein2016local,seripienlert2010dropouts,tooprakai2016simulations}. 
Flux rope structures, including smaller secondary ``islands,'' may also act as sites of particle energization \citep{Ambrosiano88,Drake06,KhabarovaEA16,MalandrakiEA19}.

The association of discontinuities with flux tube boundaries is often observed in numerical simulations of MHD and plasma turbulence  \citep{MattMont80,ServidioEA10-recon,HaggertyEA17}. On the basis of this evidence, it is clear that sharp discontinuities are frequently encountered at peripheral and internal boundaries of flux tubes. However this is not a universal property of flux ropes, and such sharp boundaries are not expected to always be present or to completely envelop flux tubes since they are often the product of the ongoing {\it interaction} of adjacent tubes, which may be a time-dependent or even sporadic process. 

In this paper, we have suggested the cooperative use of two relatively simple methods as an approach to identify flux tubes and coherent current structures that may form at their boundaries. We have chosen to employ the PVI method for identification of discontinuities \citep{greco2009statistical, greco2018partial}, and to use it in conjunction with a real-space method for systematically identifying magnetic helicity concentrations, which may be recognized as signatures of helical flux ropes.  Neither of these methods provides an absolute identification nor as complete a taxonomy as would be provided by other available methodologies. For example, traditional discontinuity identification methods \citep{tsurutani1979interplanetary,burlaga1969tangential} and their extensions \citep{VasquezEA07-disconts,Bruno2001identifying,Neugebauer06} allow for more complete classifications of classical MHD and plasma discontinuities. Similarly, various more complete techniques have been developed for identifying or visualizing the flux ropes themselves  \citep{KleinBurlaga82,Borovsky2008flux}.  A particularly elegant class of methods is the Grad Shafronov reconstruction approach \citep{Hu2017GSreview}. The real-space helicity identification approach provides a less complete picture of flux tube structure than these.

The present combination has the advantage of being relative free of assumptions concerning the types of structures that are being identified. PVI is unbiased concerning discontinuity types, and readily detects TDs, RDs. shocks, etc. Likewise, the only assumption in developing the real space helicity approach is that the statistics of the fluctuations are spatially homogeneous (or, for a time series, time stationary). No assumptions about two-dimensionality  or other spatial symmetry is required, in contrast to the standard GS method. In addition, the PVI and helicity identification methods have the practical advantage in their simplicity and ease of implementation. We may conclude that the proposed pair of methods have distinct advantages in locating likely flux ropes and their likely boundaries in data streams such as typical single spacecraft solar wind data, as well as in the analysis of very large simulation datasets. This methodology may also prove useful as a first stage of analysis to locate data that is suitable for more elaborate analyses such as Grad Shafronov reconstruction.

\begin{acknowledgements}
WHM is partially supported by  the Parker Solar Probe mission through the ISOIS Theory  and Modeling team and a subcontract from Princeton University (SUB0000317). This project has received funding from the European Unions Horizon 2020 research and innovation program under grant agreement No. 776262 (AIDA, www.aida-space.eu). The PSP data used here is publicly available on NASA CDAWeb \url{https://cdaweb.gsfc.nasa.gov/index.html/}.
\end{acknowledgements}

\bibliographystyle{aa} 
\bibliography{biblio.bib} 

\begin{thebibliography}{66}
\expandafter\ifx\csname natexlab\endcsname\relax\def\natexlab#1{#1}\fi

\bibitem[{{Ambrosiano} {et~al.}(1988){Ambrosiano}, {Matthaeus}, {Goldstein}, \&
  {Plante}}]{Ambrosiano88}
{Ambrosiano}, J., {Matthaeus}, W.~H., {Goldstein}, M.~L., \& {Plante}, D. 1988,
  Journal of Geophysical Research, 93, 14383

\bibitem[{Bale {et~al.}(2016)Bale, Goetz, Harvey, Turin, Bonnell, De~Wit,
  Ergun, MacDowall, Pulupa, Andr{\'e}, {et~al.}}]{BaleEA16}
Bale, S., Goetz, K., Harvey, P., {et~al.} 2016, Space science reviews, 204, 49

\bibitem[{{Bale} {et~al.}(2019){Bale}, {Badman}, {Bonnell}, {Bowen}, {Burgess},
  {Case}, {Cattell}, {Chandran}, {Chaston}, {Chen}, {Drake}, {de Wit},
  {Eastwood}, {Ergun}, {Farrell}, {Fong}, {Goetz}, {Goldstein}, {Goodrich},
  {Harvey}, {Horbury}, {Howes}, {Kasper}, {Kellogg}, {Klimchuk}, {Korreck},
  {Krasnoselskikh}, {Krucker}, {Laker}, {Larson}, {MacDowall}, {Maksimovic},
  {Malaspina}, {Martinez-Oliveros}, {McComas}, {Meyer-Vernet}, {Moncuquet},
  {Mozer}, {Phan}, {Pulupa}, {Raouafi}, {Salem}, {Stansby}, {Stevens}, {Szabo},
  {Velli}, {Woolley}, \& {Wygant}}]{BaleEA19}
{Bale}, S.~D., {Badman}, S.~T., {Bonnell}, J.~W., {et~al.} 2019, \nat, 576, 237

\bibitem[{{Borovsky}(2008)}]{Borovsky2008flux}
{Borovsky}, J.~E. 2008, Journal of Geophysical Research (Space Physics), 113,
  A08110

\bibitem[{{Bothmer} \& {Schwenn}(1998)}]{Bothmer98structure}
{Bothmer}, V. \& {Schwenn}, R. 1998, Annales Geophysicae, 16, 1

\bibitem[{Bruno {et~al.}(1999)Bruno, Bavassano, Bianchini, Pietropaolo,
  Villante, Carbone, \& Veltri}]{bruno1999solar}
Bruno, R., Bavassano, B., Bianchini, L., {et~al.} 1999, in Magnetic Fields and
  Solar Processes, Vol. 448, 1147

\bibitem[{{Bruno} {et~al.}(2001){Bruno}, {Carbone}, {Veltri}, {Pietropaolo}, \&
  {Bavassano}}]{Bruno2001identifying}
{Bruno}, R., {Carbone}, V., {Veltri}, P., {Pietropaolo}, E., \& {Bavassano}, B.
  2001, \planss, 49, 1201

\bibitem[{Burlaga(1988)}]{Burlaga88}
Burlaga, L. 1988, Journal of Geophysical Research: Space Physics, 93, 7217

\bibitem[{{Burlaga} {et~al.}(1981){Burlaga}, {Sittler}, {Mariani}, \&
  {Schwenn}}]{Burlaga81magnetic}
{Burlaga}, L., {Sittler}, E., {Mariani}, F., \& {Schwenn}, R. 1981, \jgr, 86,
  6673

\bibitem[{{Burlaga}(1969)}]{Burlaga69}
{Burlaga}, L.~F. 1969, \solphys, 7, 54

\bibitem[{{Burlaga} \& {Ness}(1969)}]{burlaga1969tangential}
{Burlaga}, L.~F. \& {Ness}, N.~F. 1969, \solphys, 9, 467

\bibitem[{{Chen} {et~al.}(2020){Chen}, {Hu}, {Zhao}, {Kasper}, {Bale},
  {Korreck}, {Case}, {Stevens}, {Bonnell}, {Goetz}, {Harvey}, {Klein},
  {Larson}, {Livi}, {MacDowall}, {Malaspina}, {Pulupa}, \&
  {Whittlesey}}]{chen2020small}
{Chen}, Y., {Hu}, Q., {Zhao}, L., {et~al.} 2020, arXiv e-prints,
  arXiv:2007.04551

\bibitem[{Chhiber {et~al.}(2020)Chhiber, Goldstein, Maruca, Chasapis,
  Matthaeus, Ruffolo, Bandyopadhyay, Parashar, Qudsi, de~Wit, Bale, Bonnell,
  Goetz, Harvey, MacDowall, Malaspina, Pulupa, Kasper, Korreck, Case, Stevens,
  Whittlesey, Larson, Livi, Velli, \& Raouafi}]{chhiber2020clustering}
Chhiber, R., Goldstein, M.~L., Maruca, B.~A., {et~al.} 2020, The Astrophysical
  Journal Supplement Series, 246, 31

\bibitem[{{Drake} {et~al.}(2006){Drake}, {Swisdak}, {Che}, \& {Shay}}]{Drake06}
{Drake}, J.~F., {Swisdak}, M., {Che}, H., \& {Shay}, M.~A. 2006, Nature, 443,
  553

\bibitem[{Farge(1992)}]{farge1992wavelet}
Farge, M. 1992, Annual review of fluid mechanics, 24, 395

\bibitem[{Ghosh {et~al.}(1993)Ghosh, Hossain, \& Matthaeus}]{GhoshEA93-cpc}
Ghosh, S., Hossain, M., \& Matthaeus, W.~H. 1993, Comp. Phys. Comm., 74, 18

\bibitem[{Gottlieb \& Orszag(1977)}]{GottliebOrszag}
Gottlieb, D. \& Orszag, S.~A. 1977, Numerical Analysis of Spectral Methods:
  Theory and Applications (SIAM)

\bibitem[{{Greco} {et~al.}(2008){Greco}, {Chuychai}, {Matthaeus}, {Servidio},
  \& {Dmitruk}}]{greco2008intermittent}
{Greco}, A., {Chuychai}, P., {Matthaeus}, W.~H., {Servidio}, S., \& {Dmitruk},
  P. 2008, \grl, 35, L19111

\bibitem[{Greco {et~al.}(2018)Greco, Matthaeus, Perri, Osman, Servidio, Wan, \&
  Dmitruk}]{greco2018partial}
Greco, A., Matthaeus, W., Perri, S., {et~al.} 2018, Space Science Reviews, 214,
  1

\bibitem[{{Greco} {et~al.}(2009){Greco}, {Matthaeus}, {Servidio}, {Chuychai},
  \& {Dmitruk}}]{greco2009statistical}
{Greco}, A., {Matthaeus}, W.~H., {Servidio}, S., {Chuychai}, P., \& {Dmitruk},
  P. 2009, The Astrophysical Journall, 691, L111

\bibitem[{{Haggerty} {et~al.}(2017){Haggerty}, {Parashar}, {Matthaeus}, {Shay},
  {Yang}, {Wan}, {Wu}, \& {Servidio}}]{HaggertyEA17}
{Haggerty}, C.~C., {Parashar}, T.~N., {Matthaeus}, W.~H., {et~al.} 2017,
  Physics of Plasmas, 24, 102308

\bibitem[{{Hu}(2017)}]{Hu2017GSreview}
{Hu}, Q. 2017, Sci.~China Earth Sciences, 60, 1466

\bibitem[{Hu \& Sonnerup(2002)}]{hu2002reconstruction}
Hu, Q. \& Sonnerup, B.~U. 2002, Journal of Geophysical Research: Space Physics,
  107, SSH

\bibitem[{{Hu} {et~al.}(2018){Hu}, {Zheng}, {Chen}, {le Roux}, \&
  {Zhao}}]{Hu18}
{Hu}, Q., {Zheng}, J., {Chen}, Y., {le Roux}, J., \& {Zhao}, L. 2018, \apjs,
  239, 12

\bibitem[{Jokipii(1966)}]{Jokipii66}
Jokipii, J. 1966, The Astrophysical Journal, 146, 480

\bibitem[{Jokipii(1973)}]{Jokipii73}
Jokipii, J. 1973, Annual Review of Astronomy and Astrophysics, 11, 1

\bibitem[{Jokipii \& Parker(1969)}]{JokipiiParker69}
Jokipii, J. \& Parker, E. 1969, The Astrophysical Journal, 155, 777

\bibitem[{Khabarova {et~al.}(2016)Khabarova, Zank, Li, Malandraki, le~Roux, \&
  Webb}]{KhabarovaEA16}
Khabarova, O.~V., Zank, G.~P., Li, G., {et~al.} 2016, The Astrophysical
  Journal, 827, 122

\bibitem[{Kittinaradorn {et~al.}(2009)Kittinaradorn, Ruffolo, \&
  Matthaeus}]{kittinaradorn2009solar}
Kittinaradorn, R., Ruffolo, D., \& Matthaeus, W. 2009, The Astrophysical
  Journal Letters, 702, L138

\bibitem[{{Klein} \& {Burlaga}(1982)}]{KleinBurlaga82}
{Klein}, L.~W. \& {Burlaga}, L.~F. 1982, \jgr, 87, 613

\bibitem[{Malandraki {et~al.}(2019)Malandraki, Khabarova, Bruno, Zank, Li,
  Jackson, Bisi, Greco, Pezzi, Matthaeus, Giannakopoulos, Servidio, Malova,
  Kislov, Effenberger, le~Roux, Chen, Hu, \& Engelbrecht}]{MalandrakiEA19}
Malandraki, O., Khabarova, O., Bruno, R., {et~al.} 2019, The Astrophysical
  Journal, 881, 116

\bibitem[{Matthaeus \& Goldstein(1982)}]{MattGold82a}
Matthaeus, W.~H. \& Goldstein, M.~L. 1982, \jgr, 87, 6011

\bibitem[{Matthaeus {et~al.}(1982)Matthaeus, Goldstein, \&
  Smith}]{matthaeus1982evaluation}
Matthaeus, W.~H., Goldstein, M.~L., \& Smith, C. 1982, Physical Review Letters,
  48, 1256

\bibitem[{{Matthaeus} \& {Lamkin}(1986)}]{Matthaeus86}
{Matthaeus}, W.~H. \& {Lamkin}, S.~L. 1986, Physics of Fluids, 29, 2513

\bibitem[{Matthaeus \& Montgomery(1980)}]{MattMont80}
Matthaeus, W.~H. \& Montgomery, D. 1980, Annals of the New York Academy of
  Sciences, 357, 203

\bibitem[{Matthaeus {et~al.}(2015)Matthaeus, Wan, Servidio, Greco, Osman,
  Oughton, \& Dmitruk}]{matthaeus2015intermittency}
Matthaeus, W.~H., Wan, M., Servidio, S., {et~al.} 2015, Philosophical
  Transactions of the Royal Society A: Mathematical, Physical and Engineering
  Sciences, 373, 20140154

\bibitem[{{McCracken} \& {Ness}(1966)}]{MccrackenNess66}
{McCracken}, K. \& {Ness}, N. 1966, Journal of Geophysical Research, 71, 3315

\bibitem[{Neugebauer(2006)}]{Neugebauer06}
Neugebauer, M. 2006, \jgr, 111, A04103, doi:10.1029/2005JA011497

\bibitem[{Osman {et~al.}(2014)Osman, Matthaeus, Gosling, Greco, Servidio, Hnat,
  Chapman, \& Phan}]{OsmanEA14}
Osman, K., Matthaeus, W., Gosling, J., {et~al.} 2014, Physical Review Letters,
  112, 215002

\bibitem[{Panchev(1971)}]{Panchev}
Panchev, S. 1971, Random Functions and Turbulence (New York: Pergammon Press)

\bibitem[{Parashar {et~al.}(2020)Parashar, Goldstein, Maruca, Matthaeus,
  Ruffolo, Bandyopadhyay, Chhiber, Chasapis, Qudsi, Vech,
  {et~al.}}]{ParasharEA20}
Parashar, T., Goldstein, M., Maruca, B., {et~al.} 2020, The Astrophysical
  Journal Supplement Series, 246, 58

\bibitem[{Pecora {et~al.}(2019{\natexlab{a}})Pecora, Greco, Hu, Servidio,
  Chasapis, \& Matthaeus}]{pecora2019single}
Pecora, F., Greco, A., Hu, Q., {et~al.} 2019{\natexlab{a}}, The Astrophysical
  Journal Letters, 881, L11

\bibitem[{Pecora {et~al.}(2019{\natexlab{b}})Pecora, Pucci, Lapenta, Burgess,
  \& Servidio}]{pecora2019statistical}
Pecora, F., Pucci, F., Lapenta, G., Burgess, D., \& Servidio, S.
  2019{\natexlab{b}}, Solar Physics, 294, 114

\bibitem[{Pecora {et~al.}(2018)Pecora, Servidio, Greco, Matthaeus, D.~Burgess,
  Carbone, , \& Veltri}]{pecora2018ion}
Pecora, F., Servidio, S., Greco, A., {et~al.} 2018, Journal of Plasma Physics,
  84, 725840601

\bibitem[{{Perri} {et~al.}(2017){Perri}, {Servidio}, {Vaivads}, \&
  {Valentini}}]{perri2017numerical}
{Perri}, S., {Servidio}, S., {Vaivads}, A., \& {Valentini}, F. 2017, The
  Astrophysical Journal Supplement Series, 231, 4

\bibitem[{{Ruffolo} {et~al.}(2003){Ruffolo}, {Matthaeus}, \&
  {Chuychai}}]{Ruffolo03}
{Ruffolo}, D., {Matthaeus}, W.~H., \& {Chuychai}, P. 2003, The Astrophysical
  Journall, 597, L169

\bibitem[{{Scolini} {et~al.}(2019){Scolini}, {Rodriguez}, {Mierla}, {Pomoell},
  \& {Poedts}}]{Scolini2019observation}
{Scolini}, C., {Rodriguez}, L., {Mierla}, M., {Pomoell}, J., \& {Poedts}, S.
  2019, \aap, 626, A122

\bibitem[{Seripienlert {et~al.}(2010)Seripienlert, Ruffolo, Matthaeus, \&
  Chuychai}]{seripienlert2010dropouts}
Seripienlert, A., Ruffolo, D., Matthaeus, W., \& Chuychai, P. 2010, The
  Astrophysical Journal, 711, 980

\bibitem[{{Servidio} {et~al.}(2011){Servidio}, {Greco}, {Matthaeus}, {Osman},
  \& {Dmitruk}}]{Servidio11}
{Servidio}, S., {Greco}, A., {Matthaeus}, W.~H., {Osman}, K.~T., \& {Dmitruk},
  P. 2011, Journal of Geophysical Research (Space Physics), 116, A09102

\bibitem[{{Servidio} {et~al.}(2009){Servidio}, {Matthaeus}, {Shay}, {Cassak},
  \& {Dmitruk}}]{servidio2009magnetic}
{Servidio}, S., {Matthaeus}, W.~H., {Shay}, M.~A., {Cassak}, P.~A., \&
  {Dmitruk}, P. 2009, Physical Review Letters, 102, 115003

\bibitem[{Servidio {et~al.}(2010)Servidio, Matthaeus, Shay, Dmitruk, Cassak, \&
  Wan}]{ServidioEA10-recon}
Servidio, S., Matthaeus, W.~H., Shay, M.~A., {et~al.} 2010, Physics of Plasmas,
  17

\bibitem[{Servidio {et~al.}(2014)Servidio, Matthaeus, Wan, Ruffolo, Rappazzo,
  \& Oughton}]{ServidioEA14-flux}
Servidio, S., Matthaeus, W.~H., Wan, M., {et~al.} 2014, \apj, 785, 56

\bibitem[{{Sonnerup} \& {Guo}(1996)}]{Sonnerup96}
{Sonnerup}, B.~U.~{\"O}. \& {Guo}, M. 1996, \grl, 23, 3679

\bibitem[{{Sonnerup} {et~al.}(2016){Sonnerup}, {Hasegawa}, {Denton}, \&
  {Nakamura}}]{Sonnerup16}
{Sonnerup}, B.~U.~{\"O}., {Hasegawa}, H., {Denton}, R.~E., \& {Nakamura},
  T.~K.~M. 2016, Journal of Geophysical Research (Space Physics), 121, 4279

\bibitem[{Taylor(1974)}]{Taylor74}
Taylor, J.~B. 1974, Physical Review Letters, 33, 1139

\bibitem[{Telloni {et~al.}(2012)Telloni, Bruno, D'Amicis, Pietropaolo, \&
  Carbone}]{telloni2012wavelet}
Telloni, D., Bruno, R., D'Amicis, R., Pietropaolo, E., \& Carbone, V. 2012, The
  Astrophysical Journal, 751, 19

\bibitem[{{Tessein} {et~al.}(2013){Tessein}, {Matthaeus}, {Wan}, {Osman},
  {Ruffolo}, \& {Giacalone}}]{TesseinEA13}
{Tessein}, J.~A., {Matthaeus}, W.~H., {Wan}, M., {et~al.} 2013, The
  Astrophysical Journall, 776, L8

\bibitem[{Tessein {et~al.}(2016)Tessein, Ruffolo, Matthaeus, \&
  Wan}]{tessein2016local}
Tessein, J.~A., Ruffolo, D., Matthaeus, W.~H., \& Wan, M. 2016, Geophysical
  Research Letters, 43, 3620

\bibitem[{Tooprakai {et~al.}(2007)Tooprakai, Chuychai, Minnie, Ruffolo, Bieber,
  \& Matthaeus}]{tooprakai2007temporary}
Tooprakai, P., Chuychai, P., Minnie, J., {et~al.} 2007, Geophysical research
  letters, 34

\bibitem[{Tooprakai {et~al.}(2016)Tooprakai, Seripienlert, Ruffolo, Chuychai,
  \& Matthaeus}]{tooprakai2016simulations}
Tooprakai, P., Seripienlert, A., Ruffolo, D., Chuychai, P., \& Matthaeus, W.
  2016, The Astrophysical Journal, 831, 195

\bibitem[{{Tsurutani} \& {Smith}(1979)}]{tsurutani1979interplanetary}
{Tsurutani}, B.~T. \& {Smith}, E.~J. 1979, \jgr, 84, 2773

\bibitem[{{V{\'a}sconez} {et~al.}(2015){V{\'a}sconez}, {Pucci}, {Valentini},
  {Servidio}, {Matthaeus}, \& {Malara}}]{VasconezEA15}
{V{\'a}sconez}, C.~L., {Pucci}, F., {Valentini}, F., {et~al.} 2015, \apj, 815,
  7

\bibitem[{Vasquez {et~al.}(2007)Vasquez, Abramenko, Haggerty, \&
  Smith}]{VasquezEA07-disconts}
Vasquez, B.~J., Abramenko, V.~I., Haggerty, D.~K., \& Smith, C.~W. 2007, \jgr,
  112

\bibitem[{Wan {et~al.}(2014)Wan, Rappazzo, Matthaeus, Servidio, \&
  Oughton}]{WanEA14}
Wan, M., Rappazzo, A.~F., Matthaeus, W.~H., Servidio, S., \& Oughton, S. 2014,
  The Astrophysical Journal, 797, 63

\bibitem[{Woltjer(1958)}]{Woltjer58b}
Woltjer, L. 1958, Proceedings of the National Academy of Sciences of the United
  States of America, 44, 489

\bibitem[{Zhao {et~al.}(2020)Zhao, Zank, Adhikari, Hu, Kasper, Bale, Korreck,
  Case, Stevens, Bonnell, {et~al.}}]{zhao2020identification}
Zhao, L.-L., Zank, G., Adhikari, L., {et~al.} 2020, The Astrophysical Journal
  Supplement Series, 246, 26

\end{thebibliography}
\end{document}